\documentclass[aps,nofootinbib,prd,eqsecnum,showpacs,showkeys,preprintnumbers,altaffilletter,twocolumn]{revtex4-1}

\usepackage{graphicx}
\usepackage{euscript,amssymb}
\usepackage{epsfig}
\usepackage{xcolor}
\usepackage{amsfonts}
\usepackage{amsmath}
\usepackage{amssymb}
\usepackage{fancyhdr}
\usepackage{esint}
\usepackage{enumitem}
\usepackage{booktabs}
\usepackage{array,multirow,graphicx}

\makeatother

\begin{document}

\title{More on the holographic Ricci dark energy model: smoothing Rips
through interaction effects?}

\author{Mariam Bouhmadi-L\'{o}pez $^{1,2}$}
\email{mariam.bouhmadi@ehu.eus}

\author{Ahmed Errahmani $^{3}$}
\email{ahmederrahmani1@yahoo.fr}

\author{Taoufik Ouali $^{3}$}
\email{ouali1962@gmail.com}

\author{Yaser Tavakoli $^{4,5,6}$}
\email{yaser.tavakoli@fuw.edu.pl}

\date{\today}

\affiliation{
{\mbox{${}^1$Department of Theoretical Physics University of the Basque Country UPV/EHU. P.O. Box 644, 48080 Bilbao, Spain}}\\
{\mbox{${}^2$IKERBASQUE, Basque Foundation for Science, 48011, Bilbao, Spain}}\\
{\mbox{${}^3$Laboratory of Physics of Matter and Radiation, Mohammed University, BP 717, Oujda, Morocco}}\\
{\mbox{${}^4$Faculty of Physics, University of Warsaw, Pasteura 5, 02-093 Warsaw, Poland}}
{\mbox{${}^5$School of Engineering Science, College of Engineering, University of Tehran, 11155-4563 Tehran, Iran}}
{\mbox{${}^6$School of Astronomy, Institute for Research in Fundamental Sciences (IPM), 19395-5531 Tehran, Iran}}
}

\begin{abstract}

The background cosmological dynamics of the late Universe is analysed on the framework of a dark energy model described by an holographic Ricci dark energy component. Several kind of interactions between the dark energy and the dark matter components are considered herein. We solve the background cosmological dynamics for the different choices of interactions with the aim to analyse not only the current evolution of the universe but also its asymptotic behaviour and, in particular, possible future singularities removal. We show that in most of the cases, the Big Rip singularity, a finger print of this model in absence of an interaction between the dark sectors,  is substituted by a de Sitter or a Minkowski state. Most importantly, we found two new  {\textit{ future}}  bouncing solutions leading to two possible asymptotic behaviours, we named Little Bang and Little Sibling of the Big Bang. At a Little Bang, as the size of the universe shrinks to zero in an infinite cosmic time, the Hubble rate and its cosmic time derivative blow up. In addition, at a Little sibling of the Big Bang, as the size of the universe shrinks to zero in an infinite cosmic time, the Hubble rate blows up but its cosmic time derivative is finite. These two abrupt events can happen as well in the past.

\end{abstract}

\maketitle



\section{Introduction}

\label{intro}

Several astrophysical observations (cf. for example supernovae type Ia \cite%
{Perlmutter:1998np}, cosmic microwave background \cite{Komatsu:2010fb},
large scale structure \cite{Tegmark2004}, etc.) have confirmed that the
universe is undergoing a state of accelerated expansion if homogeneity and
isotropy are assumed on large scales. In addition, those experiments indicate
that the matter content (i.e., the total mass-energy) of the universe,
leading this accelerated expansion, must contain an exotic energy which is
characterised with a sufficiently negative pressure. Dark energy (DE) is the
most accepted hypothesis to explain the current observations, and
constitutes roughly $70\%$ of the total matter content of the universe.
However, so far there is no clear understanding of the true fundamental
nature of DE. Indeed, the mysterious nature of dark energy is still among
the long-standing problems in theoretical physics.

There are many dynamical models trying to explain the nature of DE \cite{Capozziello:2010zz,AmendolaTsujikawa}. Among them, there is an
attractive model which is inspired on the holographic principle rooted on
quantum gravity \cite{'tHooft:1993gx,Susskind:1994vu,Cohen:1998zx}; it is
the so called holographic dark energy \cite{Li:2004rb,Hsu:2004ri}. We next
summarise the ideas behind this model. A well known fact is that the entropy
of a given system with finite volume, $L^3$, has an upper bound which is not
proportional to its volume, but rather to its surface area, $L^2$ \cite%
{Bekenstein:1980jp,GonzalezDiaz:1983yf}. In addition, for an effective
quantum field theory with a given ultra-violet (UV) cutoff, $M_{\text{UV}}$,
the entropy of that system scales as $L^3M^3_{\text{UV}}$. Consequently,
there is always a scale or a length where the quantum field theory with UV
cutoff is expected to fail. This is expected to happen for large volumes or
lengths. To solve this problem a link between UV and infrared (IR) cutoffs
was proposed in \cite{Cohen:1998zx}: $L^3M^4_{\text{UV}}\lesssim L M_{%
\mathrm{P}}^2.$ By this mean the validity of the quantum field theory within
this regime is assured. When the inequality is saturated, we can define an
energy density which is inversely proportional to the square of the
characteristic length of the system. These ideas have been applied to the
universe giving rise to what is known as the holographic dark energy
scenario \cite{Li:2004rb}. The appealing holographic Ricci dark energy (HRDE)
model consists in taking the square of the length characterising the
Universe, $L$, as the inverse of the Ricci scalar curvature \cite%
{Holographich} (see also \cite%
{Nojiri:2005pu,Granda:2008dk,Duran:2010ky,Chimento:2011pk,Chimento:2012fh,Chimento:2013se,Chimento:2013qja}).

It has been proven that the HRDE is suitable to describe the current
acceleration of the universe as shown in \cite%
{Feng:2008rs,Holographich,BouhmadiLopez:2013pn,Ghaffari:2015foa,Suwa:2014pia}%
. It has been equally shown that this model might induce a big rip (BR)
singularity \cite{Holographich}; i.e., the scale factor, the Hubble
parameter and its first cosmic time derivative reach very large values in a
finite future cosmic time \cite%
{Starobinsky:1999yw,Caldwell:1999ew,Caldwell:2003vq}. This model has been
also constrained observationally \cite{Xu:2010gg}. More recent observational
constraints on the HRDE can be found in Refs.~\cite%
{Chimento:2013se,Chimento:2013qja,Suwa:2014pia,Feng:2016djj,Fu:2011ab} (cf.
Ref. \cite{Zhongxu:2017} for an extended list of references on the HRDE
scenario).

On this paper, we intend to see how the BR\footnote{%
It is worth mentioning that the fate of the BR in the HRDE within a quantum
cosmological formalism was analysed in \cite{Albarran} and proven to be
harmless once appropriate boundary conditions on the wave function of the
universe are imposed.} present on the HRDE can be removed or appeased by the inclusion
of interactions between cold dark matter (CDM) and the HRDE. An interaction
on the dark sector and within the HRDE model has been previously analysed in
{\cite{Chimento:2012fh,Chimento:2013se,Chimento:2013qja,Li:2009zs,Suwa,
Fu:2011ab,Li:2014eba}} where the main goal of these
papers was to study the adequacy (from an observational point of view) of
these models to describe the late time acceleration rather than analysing
the asymptotic behaviour of the universe. On this work, we will carry a
thorough analytical analysis of the HRDE when a CDM and DE, given through
the HRDE, are interacting. The goal of this work is to identify those
interactions that are able to remove or smooth the BR. It is worth
mentioning that an interaction on the dark sector has been favoured
observationally \cite{Salvatelli:2014zta,Abdalla:2014cla} and has been shown
to be extremely helpful to mitigate the coincidence problem (cf. the recent
review \cite{Wang:2016lxa} on this topic).

The paper is organised as follows. In section II, we review the general setup of the HRDE model in a Friedmann-Lema\^{i}tre-Robertson-Walker (FLRW) background in presence of an interaction term between DM and DE. In section III, we carry out a careful analysis of the asymptotic behaviour of the universe in this framework. Finally, in section IV, we present our conclusions.

\section{General setup}

\label{model}

We consider a spatially flat FLRW
universe, filled with matter with the energy density $\rho$, whose evolution
is described by the Friedmann equation:
\begin{equation}
3M_{\text{P}}^{2}H^{2}\ =\ \rho ,  \label{friedmann1}
\end{equation}%
where $M_{\text{P}}$ is the reduced Planck mass. We assume that the total
energy density, $\rho$, of the cosmic fluid is described through a CDM
component with the energy density $\rho _{\mathrm{m}}$ and a
HRDE component $\rho_{\mathrm{H}}$. The HRDE density is defined as \cite%
{Holographich}
\begin{equation}
\rho_{\mathrm{H}}\ =\ 3\beta M_{\text{P}}^{2}\left(\frac{1}{2}\frac{dH^{2}}{%
dx}+2H^{2}\right),  \label{HRDE}
\end{equation}%
where $x \equiv \ln (a/a_0)$ and $\beta$ is a positive dimensionless
parameter that measures the strength of the holographic component. From now on a zero subindex stands for quantities evaluated at present.

It is convenient to rewrite the Friedmann equation ({\ref{friedmann1}) in
terms of the dimensionless energy densities:
\begin{align}
\Omega _{\mathrm{m}}\ &=\ \frac{\rho _{\mathrm{m}}}{3M_{\text{P}%
}^{2}H_{0}^{2}},  \notag \\
\Omega _{\mathrm{H}}\ & =\ \frac{\rho _{\mathrm{H}}}{3M_{\text{P}%
}^{2}H_{0}^{2}} =\beta \left(\dfrac{1}{2}\dfrac{dE^{2}}{dx}+2E^{2}\right),
\label{OmegaH}
\end{align}
where $E(z)=H/H_{0}$. Therefore, Eq. ({\ref{friedmann1}) becomes
\begin{equation}
E^{2}\ =\ \Omega_{\mathrm{m}}+\Omega_{\mathrm{H}}.  \label{friedmann2}
\end{equation}%
}} 
This equation constrains the cosmological parameters of the model at present
time, $x=0$, as
\begin{equation}
1=\Omega _{\mathrm{m}_{0}}+\Omega _{\mathrm{H}_{0}},  \label{friedmann3}
\end{equation}
in which the present value of the dimensionless Hubble rate $E$ and its
derivative with respect to $x$ are governed by the following equations:
\begin{equation}
\left\{
\begin{array}{c}
E(x=0)=1\text{, \ \ \ \ \ \ \ \ \ \ \ \ \ } \\
dE/dx|_{x=0}=-2+\frac{\Omega _{\mathrm{H}_{0}}}{\beta }\ .%
\end{array}
\right.  \label{E'0}
\end{equation}
In addition, at present time, the deceleration parameter $q=-(1+\frac{1}{E}%
dE/dx)$ reads,
\begin{equation}
q_{0}=1-\frac{\Omega _{\mathrm{H}_{0}}}{\beta }\ ,  \label{q0}
\end{equation}
which must be negative (i.e., $q_{0}<0$) because the universe is
accelerating currently. So that, the holographic parameter is constrained
through the inequality 
\begin{equation}
0<\beta <\Omega _{\mathrm{H}_{0}}.  \label{beta}
\end{equation}

A characteristic of the HRDE model is that the energy density of any matter
component is self-conserved. However, we will further assume in this model
an interaction between the energy densities of CDM and the HRDE components.
Therefore, the corresponding conservation equations read
\begin{align}
\dot{\rho}_{\mathrm{H}} + 3H(1+\omega_{\mathrm{H}})\rho_{\mathrm{H}}\ &=\
-Q\ ,  \label{EOSH} \\
\dot{\rho}_{\mathrm{m}} + 3H\rho_{\mathrm{m}}\ &=\ Q,  \label{EOSm}
\end{align}
where the function $Q$ denotes the interaction between the energy density of
CDM and the holographic dark energy density components. Furthermore,
positive $Q$ represents energy transfer from CDM to DE, and vice versa for
negative $Q$.

\subsection{General equations in the presence of interaction}

It is expected physically and also from Eqs. (\ref{EOSH}) and (\ref{EOSm})
that the interaction is defined through the energy densities involved in the
system, in particular $Q$ should be a function of the energy densities of
CDM and HRDE components multiplied by a quantity with the unit of inverse of
time. For convenience, we choose the Hubble rate $H$ as the characteristic
magnitude with units of inverse of time, and hence ${Q=Q(H\rho_{\mathrm{m}%
},H\rho_{\mathrm{H}}, H\rho_{\mathrm{c}})}$ {where $\rho_{\mathrm{c}}=3M_P^2H_0^2$ is the critical energy density}. Since the value of the interaction parameter $Q$ is
small (cf. Ref. \cite{Wang:2016lxa}), a power law expansion of $Q$ in terms
of the energy densities of the system is doable. The first order terms of
this interaction; i.e. linear interaction, corresponds to
\begin{equation}
Q\ \simeq\ \lambda_{\mathrm{m}}H\rho_{\mathrm{m}}+\lambda_{\mathrm{H}}H\rho_{%
\mathrm{H}}+{\lambda_{\mathrm{c}}H\rho_{\mathrm{c}}}\ ,  \label{Q}
\end{equation}
where $\lambda_{\mathrm{m}}$, $\lambda_{\mathrm{H}}$ and $\lambda_{\mathrm{c}}$ are constants.

Substituting Eq. (\ref{Q}) in Eq. (\ref{EOSm}), and replacing the energy
densities in terms of dimensionless parameters, the conservation equation (%
\ref{EOSm}) can be written as
\begin{equation}
\dfrac{d\Omega_{\mathrm{m}}}{dx}=-\Omega_{\mathrm{m}}\left(3-\lambda_{%
\mathrm{m}}+\lambda_{\mathrm{H}}\right) +\lambda_{\mathrm{H}}E^{2}+{\lambda_{\mathrm{c}}}.
\label{variation of Omega m}
\end{equation}
On the other hand, the Friedmann equation (\ref{friedmann2}) can be
rewritten as a differential equation of the dimensionless Hubble rate $E$
with respect to $x$ as
\begin{equation}
\dfrac{dE^{2}}{dx}=2\left(\frac{1}{\beta}-2\right)E^{2}-\frac{2}{\beta}%
\Omega _{\mathrm{m}}.  \label{variation of E}
\end{equation}
Notice that the equations (\ref{variation of
Omega m}) and  (\ref{variation of E}) form a system of coupled equations. However, by differentiating
both sides of Eq.~(\ref{variation of E}) with respect to $x$, and using Eq.~(%
\ref{variation of Omega m}), we obtain a second order differential equation
for $E^2$:
\begin{align}
\frac{d^{2}E^{2}}{dx^{2}}\ =\ & 
-{ 2\frac{\lambda_{\mathrm{c}}}{\beta}} + 2\left[\frac{3}{\beta}- 6 
+ \left(2-\frac{1}{\beta}\right)\lambda_{%
\mathrm{m}} - 2\lambda_{\mathrm{H}}\right]E^{2}
 \notag \\
&\  + \left(\frac{2}{\beta}+\lambda_{\mathrm{m}%
}-\lambda_{\mathrm{H}}-7\right)\frac{dE^{2}}{dx} ~\cdot  
\label{E second}
\end{align}
The total conservation law for the system is given by $\dot{\rho}%
+3H(\rho+\omega_{\mathrm{H}}\rho_{\mathrm{H}})=0$ where the total energy
density is given by $\rho=3M_{\text{P}}^2H_0^2E^2$. The equation of state
(EoS) for the HRDE is
\begin{align}
\omega_{\mathrm{H}}\ &=\ -\frac{1}{\Omega_{\mathrm{H}}}\left(E^2+\frac{1}{3}
\frac{dE^2}{dx}\right).  \label{EoS1}
\end{align}
This EoS can be rewritten by using Eqs.~(\ref{OmegaH})-(\ref{friedmann2})  as
\begin{align}
\omega_{\mathrm{H}}\ &=\ \frac{1}{3}\left(\frac{\Omega_{\mathrm{m}}}{\Omega_{%
\mathrm{H}}}+1-\frac{2}{\beta }\right)  \notag \\
\ & =\ \frac{1}{3}\left(\frac{E^{2}}{\Omega_{\mathrm{H}}}-\frac{2}{\beta}%
\right).  \label{EoS2}
\end{align}
Notice that Eq. (\ref{EoS2}) is valid for any HRDE model where the total
energy density of the universe is conserved.

In order to study the behaviour of the universe within the context of the
HRDE, we will analyse the solutions of the differential equation (\ref{E
second}) which depend on $\lambda_{\mathrm{m}}$, $\lambda_{\mathrm{H}}$, 
{$\lambda_{\mathrm{c}}$} and $%
\beta$ parameters. We remind that we are interested in analysing the phase
space of the parameters of the model for which the universe has a smooth
future behaviour.


\subsection{Expected behaviour: general solutions}
\label{sol-general}

The general solution of the equation (\ref{E second}) can be written as:
\begin{equation}
E^{2}(x)\ =\ \mathbf{A}_{+}e^{\sigma _{+}x}+\mathbf{A}_{-}e^{\sigma _{-}x} + {\Lambda_{\mathrm{c}}}~,
\label{solution}
\end{equation}%
with $\mathbf{A}_{\pm }$ being constants,  
\begin{eqnarray}
{\Lambda_{\mathrm{c}}\ :=\ \frac{\lambda_{\mathrm{c}}}{3-6\beta+(2\beta-1)\lambda_{\mathrm{m}}-2\lambda_{\mathrm{H}}\beta}} ~,
\end{eqnarray}
and
\begin{eqnarray}
\sigma _{\pm }\ &:=&\ \sigma _{0}\pm \frac{\sqrt{\Delta }}{2\beta },  \notag
\\
\mathrm{where} \quad \sigma _{0}\ &=&\ \frac{1}{2\beta }\big[2+(\lambda _{%
\mathrm{m}}-\lambda _{\mathrm{H}}-7)\beta \big]. \quad \quad  \label{sigma}
\end{eqnarray}%
The parameter $\Delta $ reads
\begin{align}
\Delta = \left[\Big(\lambda_{\mathrm{m}} + \lambda_{\mathrm{H}} + 1 -\frac{2%
}{\beta}\Big)^2 - 4\lambda_{\mathrm{H}}(\lambda_{\mathrm{m}}+1)\right]%
\beta^2.  \label{delta}
\end{align}
On the other hand, for the case in which $\Delta =0$ and $\sigma_0\neq 0$,
the solution for $E^{2} $ reads
\begin{equation}
E^{2}(x)\ =\ (\mathbf{A}_{0}+\mathbf{A}_{1}x)e^{\sigma _{0}x}+{\Lambda_{\mathrm{c}}}~,
\label{solution2}
\end{equation}%
where $\mathbf{A}_{0}$ and $\mathbf{A}_{1}$ are constants. 
The Hubble rate $E^{2}$ at the present time must satisfy the conditions of
Eq. (\ref{E'0}), i.e., the constants $\mathbf{A}_{\pm }$, $\mathbf{A}_{0}$
and $\mathbf{A}_{1}$ can be expressed in terms of $q_{0}$, Eq.~(\ref{q0}),
as 
\begin{eqnarray}
\mathbf{A}_{\pm }\ & =& \ \pm \frac{2(1+q_{0})+{\sigma _{\mp}(1-\Lambda_{\mathrm{c}})}}{\sigma_{-}-\sigma _{+}}\ ,  \label{A+-} \\
\mathbf{A}_{0}\ & =& \ 1-{\Lambda_{\mathrm{c}}}~,  \label{A} \\
\mathbf{A}_{1}\ & =& \ -2(1+q_{0})+\sigma _{0}({\Lambda_{\mathrm{c}}}-1) \ .  \label{A1}
\end{eqnarray}

Furthermore, by substituting the solutions (\ref{solution}) and (\ref{solution2}) 
in the second equation of (\ref{OmegaH}), the dimensionless
HRDE densities read, respectively
\begin{eqnarray}
\Omega _{\mathrm{H}}\ &=&\ \frac{\beta \mathbf{A}_{+}}{2}\left(\sigma
_{+}+4\right) e^{\sigma _{+}x} \nonumber \\
&& \quad  + \frac{\beta \mathbf{A}_{-}}{2}\left( \sigma
_{-}+4\right) e^{\sigma _{-}x}+{2\beta\Lambda_{\mathrm{c}}}~,  \quad\quad
\label{sol HRDE}
\end{eqnarray}
and
\begin{eqnarray}
\Omega _{\mathrm{H}} = \frac{\beta}{2}\big[ (4+\sigma_0)(A_0+A_1x) + A_1 
\big] e^{\sigma _0x}+2\beta\Lambda_{\mathrm{c}}.
\quad \quad   
\label{sol2 HRDE}
\end{eqnarray}

In the rest of this paper, we will analyse the solutions of Eqs.~(\ref%
{solution}) and (\ref{solution2}), and we will discuss their asymptotic
behaviours, which depend on the choices of the holographic parameter $\beta$
and the interaction constants $\lambda _{\mathrm{m}}$, $\lambda _{%
\mathrm{H}}$ and ${\lambda_{\mathrm{c}}}$. We will show that, in the far future, the fate of the universe
may end up in one of the following states: a BR singularity \cite%
{Starobinsky:1999yw,Caldwell:1999ew,Caldwell:2003vq}; a Minkowskian or a de
Sitter behaviour; a little sibling of the BR (LSBR) \cite%
{IJMPD24}; we will as well show for the first time the
presence of two other possible asymptotic behaviours corresponding to what
we named the little bang (LB) and the little sibling of the big bang
(LSBB)(cf. sections \ref{subsec.B} and \ref{Sing-model3}).


\section{Asymptotic behaviour and interaction effects}
\label{Sing-model}

In this section, we will analyse the asymptotic behaviour of a FLRW filled
with an HRDE fluid interacting with CDM within the model introduced on the
previous section. In particular, we will analyse potential future
singularities that might appear on the model by considering different
interaction functions $Q\neq0$. First of all, we will start with a brief
review of the standard HRDE model in {the} absence of interactions.


\subsection{Standard HRDE model: $Q=0$}
\label{Sing-model1}

For vanishing parameters $\lambda_{\mathrm{m}}$, $\lambda_{\mathrm{H}}$ and 
{$\lambda_{\mathrm{c}}$},
in Eq. (\ref{Q}), there is no interaction (i.e., $Q=0$) and the standard
HRDE model is recovered \cite{Holographich}. 
Then, by setting 
$\lambda_{\mathrm{m}}=\lambda_{\mathrm{H}}={\lambda_{\mathrm{c}}}=0$ 
in solution (\ref{solution}) we obtain the dimensionless Hubble rate $E(x)$ (for the case
$\beta\neq 2$) as (cf. Ref \cite{Holographich})
\begin{eqnarray}
E^{2}(x)\ &=&\ \frac{2q_{0}-1}{\beta-2}\beta \exp\left[\left(\frac{2}{\beta}-4\right)x\right]  \notag \\
&& \ \ \ -\frac{2(q_0-1)\beta+2}{\beta-2}\exp(-3x).
\label{E^2 non interacting}
\end{eqnarray}
The second term in Eq.~(\ref{E^2 non interacting}) vanishes as $%
x\rightarrow+\infty$, and the Hubble rate is governed only by the first
term. This indicates that, in the far future, the HRDE energy density
mimicking matter (second term in Eq. (\ref{E^2 non interacting})) is
practically zero, and the universe converges asymptotically to a universe
filled with the dominant HRDE component. Therefore, the properties of the
solution (\ref{E^2 non interacting}) depend on the different ranges of the
holographic parameter $\beta$. Let us summarise those behaviours as follows:
\begin{enumerate}
\item { For $\beta>\frac{1}{2}$ ($\beta\neq2$), the
HRDE density tends to  zero and the universe heads to a Minkowski state in the
far future.}
\item {For $\beta=\frac{1}{2}$, the universe tends to a
\emph{de Sitter} state in the far future.} 
\item {If $0<\beta <\frac{1}{2}$, the HRDE is dominant at
late time. Using Eq.~(\ref{EoS2}), the equation of state for HRDE energy can
be written as}
\begin{equation}
\omega _{\mathrm{H}}=\frac{1}{3}\left( 1-\frac{2}{\beta }\right) ,
\end{equation}%
which is always smaller than $-1$. Therefore, the HRDE component behaves as
phantom-like matter at late time. By integrating Eq.~(\ref{E^2 non
interacting}), we obtain the evolution of the scale factor $a(t)$:
\begin{align}
& a(t)=\Big[ CH_{0}(t-t_{0})+1\Big] ^{\frac{\beta }{2\beta -1}}\ ,
\label{scale-sHRDE}
\end{align}
where
\begin{align}
\quad C:=\frac{2\beta -1}{\sqrt{\beta }}\sqrt{\frac{1-2q_{0}}{2-\beta}}.
\end{align}
Notice that, we have set $a(t_{0})=1$ as the value of the scale factor at
present. On the other hand, from Eq.~(\ref{sol HRDE}), { we find that the
dimensionless energy density of the HRDE increases negatively with
time (since $\beta <\frac{1}{2}$). Then, at a finite time, namely $t_{\mathrm{BR}}$,
the scale factor (Eq. (\ref{scale-sHRDE})), the Hubble parameter and its
cosmic time derivative blow up at $t_{\mathrm{BR}}$. Therefore, the universe
hits a BR singularity at $t_{\mathrm{BR}}$. 
In fact, it can be seen that $t_{\mathrm{BR}}$ is a finite time and depends on the
holographic parameter $\beta$.} 
\end{enumerate}
{Finally, for  the case
 $\beta=2$ (which is not included in Eq.~(\ref{E^2 non interacting})),
 since $q_0<\frac{1}{2}$,  the universe tends to a {\em Minkowski} state in the far future.}

On the one hand, the latest observational data (e.g., Planck results \cite{Komatsu:2010fb}), implies
that $\Omega _{m}\sim 0.308$  and  $H_0=67.8~{\rm km}\cdot {\rm s}^{-1}\cdot {\rm Mpc}^{-1}$, thus, $q_0=-0.538$ (for a $\Lambda$CDM universe).  Then, by considering the condition~(\ref{q0}),
we can estimate the holographic parameter $\beta $ to be of the
order $\beta \sim 0.448$. By setting $t_{0}=H_{0}^{-1}\sim 14.422$
Gyrs, we find  that the BR  would take place when $t_{\mathrm{BR}}=94.675$ Gyrs.
On the other hand, we have seen that for $\beta\geq\frac{1}{2}$ there would be no abrupt events
or singularities at late time.
Given that we are interested in analysing DE singularities in this model, which are observationally favoured, from now on we will disregard values of $\beta$ such that $\beta \geq \frac{1}{2}$.

\subsection{The solution with  $\Delta=\protect\sigma_0=0$}
\label{subsec.B}

Let us now consider two particular classes of solutions 
for Eq.~(\ref{E second}). For one of these solutions,
the right hand side of Eq.~(\ref{E second}) vanishes for any value of the scale factor.
This corresponds to the solution (\ref{solution2}) for $\Delta=0$, $\sigma_0=0$ and ${\lambda_{\mathrm{c}}=0}$.
{Consequently, we have
\begin{eqnarray}
\lambda_{\mathrm{H}} = \frac{2}{\beta}(2\beta-1)^2~, \quad
\lambda_{\mathrm{m}} =  8\beta-1~, \quad 
\lambda_\mathrm{c} = 0. \quad \quad  \label{caseSolB}
\end{eqnarray}
Then, the corresponding solution for $E(x)$ in
this case can be written as}
\begin{equation}
E^{2}(x)\ =\ Ax+B,  \label{LR1}
\end{equation}
where $A$ and $B$ are constants. 
By applying the condition (\ref{E'0}), and using Eq. (\ref{q0}), we obtain
\begin{equation}
B=1,\quad \quad A = 2\left(\frac{\Omega_{\mathrm{H}_0}}{\beta}-2\right) =
-2(1+q_0) .  \label{AB1}
\end{equation}
Since we expect $q_0>-1$ (this can be proven from a rough estimation based
on the $\Lambda$CDM model), from the right hand side of Eq.~(\ref{AB1}), we
expect $A$ to be always negative. 
Positiveness of $E^2$ in Eq.~(\ref{LR1}) implies that, $x$ always lies in
the range $-\infty<x\leq \frac{B}{|A|}$.

By using Eq.~(\ref{LR1}) in the relation $dx/dt=H=H_{0}E$ and integrating
both sides, we obtain, for negative values of $A$, a relation for the time
dependence of $x$, as
\begin{equation}
x(t)=-\frac{|A|H_{0}^{2}}{4}\left( t-t_{0}-\frac{2}{|A|H_{0}}\right) ^{2}+%
\frac{1}{|A|}\ ,  \label{x-2}
\end{equation}%
where $t_{0}$ denotes the present time for which $x(t_{0})=0$. Notice that,
we have set $B=1$ in the above relation. By taking the time derivative of $x$%
, and replacing it on the left hand side of equation $dx/dt=H_{0}E$, we
obtain the dimensionless Hubble parameter
\begin{equation}
E(t)\ =\frac{A}{2}H_{0}(t-t_{0})+1\ .  \label{HR1}
\end{equation}%
Consequently, the time derivative of the Hubble rate (\ref{HR1}) reads
\begin{equation}
\dot{E}(t)\ =\frac{A}{2}H_{0},  \label{HR2}
\end{equation}%
which is constant during the evolution of the universe.

Since $-\infty <x\leq \frac{1}{|A|}$ in this case, at $t_{b}$ where
\begin{equation*}
t_{b}\ =\ t_{0}+\frac{2}{|A|H_{0}}~,
\end{equation*}%
{$x(t)$ reaches its upper limit $x=\frac{1}{|A|}$, where the Hubble rate
vanishes, and the first time derivative of the Hubble rate, Eq.~(\ref{HR2}),
remains constant. Thus, at this point the universe hits a bounce.
Hereafter, the universe starts to collapse
during an {\em infinite} time, and as $x\rightarrow-\infty$, for which  $a\rightarrow0$, the Hubble rate (\ref{HR1}) diverges while its time derivative (\ref{HR2}) remains finite.
This represents a new abrupt event  for the future fate of the universe
which happens at $a=0$, and  is  different  from  the big bang singularity
(for which the Hubble rate and its time derivative diverge). This abrupt event is smoother than a big bang:
we named it the ``\emph{little sibling of the big bang}" (LSBB).}

By using Eq.~(\ref{friedmann1}) and the total conservation equation $\dot{\rho}+3H(1+\omega_{\mathrm{tot}})\rho=0$, the total EoS
$\omega_{\mathrm{tot}}=p_{\mathrm{H}}/(\rho_{\mathrm{H}}+\rho_{\mathrm{m}})$ reads
\begin{eqnarray}
\omega_{\rm tot}\ &=&\ -1-\frac{1}{3E^{2}}\frac{dE^{2}}{dx} \nonumber \\ 
&=&\ -1-\frac{A}{3(Ax+1)}\  \cdot
\end{eqnarray}
In the limit  $x\rightarrow-\infty$, the total EoS $\omega_{\rm tot}$ tends to $-1$. We would like to stress that the evolution of the universe is symmetric with regards to the bounce. Therefore, the universe evolves from a LSBB to a bounce and recollapse heading back to a LSBB.

The second class of solution is the case
in which  parameters $\lambda_{\mathrm{H}}$ and $\lambda_{%
\mathrm{m}}$ satisfy the conditions~(\ref{caseSolB}) but $\lambda_{\mathrm{c}}\neq0$. 
In this case Eq.~(\ref{E second}) reduces to 
\begin{equation}
\frac{d^{2}E^{2}}{dx^{2}}\ =\ -\frac{2}{\beta}\lambda_{\mathrm{c}}\ .
\label{Eq-ext2}
\end{equation}
With respect  to the cosmic time $t$, Eq.~(\ref{Eq-ext2}) becomes 
\begin{equation}
\ddot{E}(t)-\omega^2 E(t)=0\ ,
\end{equation}
where,  $\omega^2=-\frac{\lambda_{\mathrm{c}}}{\beta}H_{0}^2$.
By imposing the conditions (\ref{E'0}) and (\ref{q0}), the dimensionless Hubble rate reads
\begin{equation}
E(t)\ =\ C_{+}\exp(\omega t)+C_{-}\exp(-\omega t)\ ,
\end{equation}
{where, from Eq.~(\ref{E'0}) at  present time $t=t_0$, we get
\begin{eqnarray}
C_\pm \ &=&\  \frac{1}{2}\Big[1 \mp \frac{H_0}{\omega}(1+q_0)\Big]\exp(\mp\omega t_0)\ .
\end{eqnarray}}
Moreover,  the scale factor is given by\footnote{We thank the referee for reminding us  this class of solutions.}
\begin{eqnarray}
a(t) = a_{0} \exp\left[\frac{H_{0}}{\omega}C_{+}e^{\omega t}-\frac{H_{0}}{\omega}C_{-}e^{-\omega t}\right]. \quad \quad 
\end{eqnarray}                                             
Here, $a_{0}$ is a constant of integration. 
Notice that, as $t\rightarrow+\infty$  the energy density (\ref{HRDE}) of the universe diverges.
Likewise, the Hubble rate blows up in this case.

In order to better understand the behaviour of the Hubble rate (\ref{Eq-ext2}),
it is convenient to write its solution in the following form:
\begin{equation}
E^{2}(x)\ =\ -\frac{\lambda _{\mathrm{c}}}{\beta }\left( x-x_{1}\right)
\left(x-x_{2}\right),  \label{friedmann2bb-1}
\end{equation}
where $x_{1}$ and $x_{2}$ are defined as
\begin{align}
x_{1}\ & :=\ -\frac{\beta}{\lambda _{\mathrm{c}}}\left((1+q_{0})+\sqrt{%
(1+q_{0})^{2} +\frac{\lambda _{\mathrm{c}}}{\beta }}\right) ,
\label{friedmann2bb-2} \\
x_{2}\ & :=\ -\frac{\beta}{\lambda _{\mathrm{c}}}\left((1+q_{0})-\sqrt{%
(1+q_{0})^{2} +\frac{\lambda _{\mathrm{c}}}{\beta }}\right) .
\label{friedmann2bb-3}
\end{align}
{{For $\lambda _{\mathrm{c}}<0$, 
the only physically interesting situations takes place when $x$ 
belongs to the range $-\infty <x<x_{2}$: 
{{this  describes a universe that starts its evolution from
an infinite time in the past (where $x\rightarrow-\infty$ and $a\rightarrow0$), 
at which the Hubble rate, $E(x)$, and  its time derivative, $\dot{E}$, diverge.
Therefore, the universe meets a new abrupt event in this case
which is similar to the big bang singularity, but occurs during  an  infinite cosmic time.
We thus name this new abrupt event as the ``{\em little  bang}" (LB).}
After this point, the universe expands till $x=x_2$ and then it bounces back to a LB.}
The other situation, $x_1\leq x$, corresponds to an expansion that starts at a bounce when $x= x_1$ and heads to a little rip, when $x\rightarrow +\infty$ and $t\rightarrow +\infty$ \cite{Ruzmaikina,Nojiri:2005sx,Nojiri:2005sr,Stefancic:2004kb,BouhmadiLopez:2005gk,Frampton:2011sp,Brevik:2011mm,Bouhmadi-Lopez:2013nma}.
Finally,  when $\lambda_{\rm c}>0$, there is a unique Lorentzian solution that interpolate between two bounces located at $x=x_1=x_2$.}}

\subsection{{ Interacting HRDE model}}
\label{Sing-model3}
{In this section, and in order to illustrate our purpose, we consider an interacting HRDE model with only the arbitrary parameters $%
\lambda _{\mathrm{H}}$ and $\lambda _{\mathrm{m}}$. }
By introducing a new quantity $r$, the ratio
between the energy densities of CDM and HRDE \cite{Pav1, Pav2}:
\begin{equation}
r\ :=\ \frac{\rho _{\mathrm{m}}}{\rho _{\mathrm{H}}}\ =\ \frac{\Omega _{%
\mathrm{m}}}{\Omega _{\mathrm{H}}}\ ,  \label{r}
\end{equation}
and using the conservation Eqs.~(\ref{EOSH}) and (\ref{EOSm}), we get
\begin{equation}
\frac{dr}{dx}\ =\ (r+1)\frac{Q}{H\rho _{\mathrm{H}}}+3r\omega _{\mathrm{H}}\ .
\label{r dot}
\end{equation}
{By substituting the interaction $Q=H(\lambda _{\mathrm{m}%
}r+\lambda _{\mathrm{H}})\rho_{\mathrm{H}}$ in 
Eq. (\ref{r dot}), we obtain
\begin{equation}
\frac{dr}{dx}\ =\ (\lambda _{\mathrm{m}}+1)r^{2}+\left(\lambda_{\mathrm{H}%
}+\lambda _{\mathrm{m}}+1-\frac{2}{\beta }\right)r 
+\lambda_{\mathrm{H}}\cdot 
\label{rII-1}
\end{equation}
The solutions for }the differential equation~(\ref{rII-1})
depend on the sign of the discriminant $\Delta$, Eq. (\ref{delta}), which can be rewritten as
\begin{equation}
\Delta\ :=\ \alpha \big(\beta -\beta_{1}\big)\big(\beta -\beta
_{2}\big),  \label{Delta}
\end{equation}
where $\alpha $, $\beta _{1}$ and $\beta _{2}$ are:
\begin{align}
\alpha\ & =\ (\lambda_{\mathrm{H}}+\lambda _{\mathrm{m}}+1)^{2}-4\lambda_{%
\mathrm{H}}(\lambda_{\mathrm{m}}+1) ~,  \label{alpha} \\
\beta _{1} \ & =\ \frac{2}{\alpha}\left[\lambda_{\mathrm{H}}+\lambda _{%
\mathrm{m}}+1 + 2\sqrt{\lambda_{\mathrm{H}}(\lambda_{\mathrm{m}}+1)}\right],
\label{beta1} \\
\beta _{2} \ & = \ \frac{2}{\alpha}\left[\lambda_{\mathrm{H}}+\lambda _{%
\mathrm{m}}+1 - 2\sqrt{\lambda_{\mathrm{H}}(\lambda_{\mathrm{m}}+1) }\right].
\label{beta2}
\end{align}
To get the physical solutions of Eq. (\ref{rII-1}), it is helpful to
distinguish three cases $\Delta>0$, $\Delta=0$ and $\Delta<0$: 

\begin{enumerate}
\item When $\Delta>0$, the parameter $\beta $ fulfils $\beta >\beta _{1}$ or $%
\beta <\beta _{2}$ (where $\beta_{2}<\beta _{1}$). In this case, the
solution of the differential equation (\ref{rII-1}) reads, 
\begin{align}
\quad \quad r(x) = -\frac{\mathbf{b}}{2\mathbf{a}} + \frac{\sqrt{\Delta}}{2\mathbf{a}{\beta}} \tanh \Big(-\frac{\sqrt{\Delta}}{2{\beta}}(x+k_1)\Big),  
\label{r-beta1}
\end{align}
where $\mathbf{a} := \lambda _{\mathrm{m}}+1$, $\mathbf{b}:=\lambda _{\mathrm{m%
}}+\lambda_{\mathrm{H}}+1-\frac{2}{\beta}$  and $k_1$ is a constant
of integration.  As $x\rightarrow \infty$, the ratio $r(x)$ in Eq. (\ref{r-beta1}) 
tends to a constant:
\begin{eqnarray}
\quad \quad   r(x\rightarrow\infty ) = -\frac{\sqrt{\Delta}+{\beta}(\lambda_{\mathrm{H}%
}+\lambda _{\mathrm{m}} +1-\frac{2}{\beta})}{2{\beta}(\lambda _{\mathrm{m}}+1)}
\cdot \ \ \quad 
\end{eqnarray}%
A physical solution implies that $r(x)$ must be always a positive valued function,
that is, $r(x\rightarrow\infty )\geq 0$; this corresponds to ${\beta}(\lambda_{\mathrm{H}%
}+\lambda _{\mathrm{m}}+1-\frac{2}{\beta})+\sqrt{\Delta}\leq 0$. Consequently,
this relation imposes a constraint on the holographic parameter $\beta <%
\frac{2}{1+\lambda_{\mathrm{H}}+\lambda _{\mathrm{m}}}$. Since $\beta _{1}$
is always larger than $\frac{2}{1+\lambda_{\mathrm{H}}+\lambda _{\mathrm{m}}}
$, the positivity of the function $\Delta$ implies that $\beta $ must be on the
range
\begin{align}
\beta<\beta _{2}<\frac{2}{1+\lambda_{\mathrm{H}}+\lambda _{\mathrm{m}}}
<\beta_1 \ ,
\end{align}
for the solution (\ref{r-beta1}) to be meaningful. 

In the far future (as $x\rightarrow +\infty $), by setting $\lambda_{\mathrm{H}}=0$
 and for the range
of the parameters satisfying $0<\beta <\frac{2}{1+\lambda_{\mathrm{m}} }$, the ratio of the
energy densities $r(x)$ vanishes ($r\rightarrow 0$)
and the budget content of the universe becomes dominated by the HRDE, as
would be expected. 
On the other hand, if $\beta \geq \frac{2}{1+\lambda_{\mathrm{m}} }$, the ratio $r(x)$
becomes negative which is not physically possible.
While  by setting $\lambda_{\mathrm{m}}=0$, in the far future, $r(x)$
approaches the value
\begin{eqnarray}
\quad \  r(x\rightarrow+\infty)\ =\ -\frac{1}{2}\Big(\lambda_{\mathrm{H}}-\frac{2}{\beta}+1+\frac{\sqrt{\Delta}%
}{\beta}\Big). \quad
\end{eqnarray}
Since $r(x)$ is always positive, the parameter $\beta$ is constrained to
fulfil $\beta<\frac{2}{1+\lambda_{\mathrm{H}}}$. 
\item \label{item-D0} For the case $\Delta=0$, the possible values of $\beta$ are
$\beta =\beta_{1}$ or $\beta =\beta _{2}$. In this case, the solution of
Eq.~(\ref{rII-1}) for $r(x)$ is given by 
\begin{align}
r(x)\ &=\ -\frac{b}{2\mathbf{a}} - \frac{1}{\mathbf{a}\left(x+k_2\right)}\ ,  \label{r-beta2}
\end{align}%
where $k_2$ is a constant of integration. On the limit $x\rightarrow +\infty
$, the solution (\ref{r-beta2}) tends to
\begin{equation}
r(x\rightarrow\infty )\ =\ -\frac{(\lambda_{\mathrm{H}}+\lambda _{\mathrm{m}}+1-\frac{2}{%
\beta })}{2(\lambda _{\mathrm{m}}+1)}\ \cdot
\end{equation}%
Again $r(x)$ must be positive, then, the holographic parameter must be on
the range $\beta <\frac{2}{1+\lambda_{\mathrm{H}}+\lambda _{\mathrm{m}}}$.
Since $\beta _{1}$ is larger than $\frac{2}{1+\lambda_{\mathrm{H}}+\lambda_{%
\mathrm{m}}}$, the only possible range for the holographic parameter in this
case is
\begin{align}
\beta =\beta _{2}<\frac{2}{1+\lambda_{\mathrm{H}}+\lambda _{\mathrm{m}}}%
<\beta_1.  \label{beta-possible}
\end{align}
\item Finally, if $\Delta<0$ then $\beta _{2}<\beta <\beta _{1}
$. The solution of Eq. (\ref{rII-1}) in this case is given as
\begin{align}
\quad \quad r(x) = -\frac{b}{2\mathbf{a}}+\frac{\sqrt{\left\vert \Delta\right\vert }}{2\mathbf{a}{\beta}}
\tan \Big( \frac{\sqrt{\left\vert \Delta\right\vert}}{2{\beta}}(x+k_3)\Big) ,
\label{r2}
\end{align}%
where $k_3$ is an integration constant. Equation (\ref{r2}) shows that when
the argument of the `tangent' term reaches the values $\pm\frac{\pi }{2}+n\pi $
where $n\in \mathbb{Z}$, the ratio $r$ diverges
with positive and negative signs, respectively. The former corresponds to a
universe filled with CDM at late time, which does not match with the
cosmological observations. In addition the latter limiting case corresponds to a negative
ratio $r(x)$ for the energy densities of the universe. Therefore, the
solution given by Eq.~(\ref{r2}) (that is, the solution provided by the case
$\Delta<0$) is not physically relevant.
\end{enumerate}
We will henceforth, study the late time behaviour of the universe predicted
by the two physical solutions corresponding to $\Delta>0$ and $\Delta=0$.

\subsubsection{The case $\Delta>0$}

{The dimensionless Hubble rate $E(x)$ can be obtained by using Eq. (\ref{solution}) in which
$\Lambda_{\rm c}=0$ and $\sigma_{\pm}$ is given by}
%
\begin{align}
\sigma_\pm\ =\ \frac{2+(\lambda_{\mathrm{m}}-\lambda_{\mathrm{H}%
}-7)\beta\pm\beta \sqrt{\Delta}}{2\beta}\ ,
\end{align}
with $\Delta\neq 0$ given by Eq. (\ref{Delta}). Moreover,  $\mathbf{A}_{\pm }$ reads
\begin{equation}
\mathbf{A}_{\pm }\ =\ \pm \frac{(3-4q_{0}+\lambda_{\mathrm{H}}-\lambda _{\mathrm{m}%
}\pm \sqrt{\Delta})\beta -2}{2\beta \sqrt{\Delta}}\ .  \label{sol-D2}
\end{equation}%
When $\Delta>0$, the Hubble rate (\ref{solution}) at late time becomes
\begin{eqnarray}
E^{2}(x)=\mathbf{A}_{+}\exp \Big[\frac{2+(\lambda _{\mathrm{m}}-\lambda_{\mathrm{H}%
}-7)\beta +\beta \sqrt{\Delta}}{2\beta }x\Big] . \quad  \quad \quad
\label{sol-D3}
\end{eqnarray}
For the two ranges of the parameter $\beta $ in which $\beta <\frac{2}{%
7+\lambda_{\mathrm{H}}-\lambda _{\mathrm{m}}}$ or $\frac{2}{7+\lambda_{%
\mathrm{H}}-\lambda _{\mathrm{m}}}<\beta <\frac{3-\lambda _{\mathrm{m}}}{%
6+2(\lambda_{\mathrm{H}}-\lambda _{\mathrm{m}})}$, the argument of the
exponential term in Eq.~(\ref{sol-D3}) is positive. In this case, the
universe will undergo a BR singularity in the far future. By integrating
Eq.~(\ref{sol-D3}), we can find the behaviour of the scale factor $a(t)$ at late times as
\begin{align}
a(t)\ =\ a_0\left[\frac{2}{2-\sigma_+H_0\sqrt{\mathbf{A}_+}(t-t_0)}\right]^{\frac{2}{%
\sigma_+}}\ .
\end{align}
This  indicates that the BR occurs at $t_{\mathrm{BR}}$ where
\begin{equation}
t_{\mathrm{BR}}\ =\ t_{0}+\frac{\left(4\beta/H_{0}\sqrt{\mathbf{A}_{+}}\right)}{%
2+(\lambda_{\mathrm{m}}-\lambda_{\mathrm{H}}-7)\beta +\beta \sqrt{\Delta}}\ ,
\end{equation}%
and the scale factor diverges as $a(t_{\mathrm{BR}})\rightarrow+\infty$. On
the other hand, for the range of the holographic parameter such that $\beta >%
\frac{3-\lambda _{\mathrm{m}}}{6+2(\lambda_{\mathrm{H}}-\lambda _{\mathrm{m}%
})}$, the argument of the exponential term is negative, the Hubble rate
vanishes and the universe converges to a Minkowski state in the far
future. In addition, for $\beta =\frac{3-\lambda _{\mathrm{m}}}{6+2(\lambda_{%
\mathrm{H}}-\lambda _{\mathrm{m}})}$ and $\beta \neq \frac{2}{7+\lambda_{%
\mathrm{H}}-\lambda _{\mathrm{m}}}$ the Hubble rate (\ref{sol-D3}) reduces
to a constant $E^{2}=\mathbf{A}_{+}=const.$, indicating that the universe approaches a
de Sitter state at late time.

{{By setting $\lambda_{\mathrm{H}}=0$,  Eq. (\ref{solution}) reduces to
\begin{eqnarray}
E^{2}(x) &=& \Big\{\beta\Big(\frac{\lambda_{\mathrm{m}} +2q_{0}-1}{\beta (1+\lambda_{\mathrm{m}} )-2}\Big) \exp\Big[\Big(\frac{2}{\beta}-(\lambda_{\mathrm{m}}+1)\Big)x\Big]  \nonumber \\
&&   + 2\Big(\frac{\beta(1-q_{0})-1}{\beta(1+\lambda_{\mathrm{m}})-2}\Big)\Big\} \exp\big[(\lambda_{\mathrm{m}}-3)x\big].  
\label{E1-b}
\end{eqnarray}
Now, we summarise the properties of the solution (\ref{E1-b}), for the case $%
0<\beta <\frac{2}{1+\lambda_{\mathrm{m}}}$: 
\begin{enumerate}
\item For the case $\lambda_{\mathrm{m}}\geq3$, the holographic parameter $\beta$ is
within the range $\beta <\frac{2}{1+\lambda_{\mathrm{m}}}\leq\frac{1}{2}$~. In this range,
the first term in Eq. (\ref{E1-b}) becomes negative; $\frac{\lambda_{\mathrm{m}}+2q_{0}-1}{%
\beta(1+\lambda_{\mathrm{m}})-2}<0$, and the second one becomes positive; $\frac{%
\beta(1-q_{0})-1}{\beta(1+\lambda_{\mathrm{m}})-2}>0$. Thus, the Hubble rate
decreases and vanishes at $x_b$ which means that the universe hits a bounce at
\begin{eqnarray}
 x_b=\frac{\beta}{2-\beta(1+\lambda_{\mathrm{m}})}\ln\left[\frac{2\big(\beta(q_{0}-1)+1\big)
}{\beta(\lambda_{\mathrm{m}}+2q_{0}-1)}\right].
\label{Bounce-1}
\end{eqnarray}
In  the far future, after the bounce ($x\rightarrow -\infty$), the dimensionless
Hubble rate vanishes and the universe tends to a Minkowski state.
\item For the case $\lambda_{\mathrm{m}} <3$, the holographic parameter $\beta $ satisfies
the condition $\beta<\frac{1}{2}<\frac{2}{1+\lambda_{\mathrm{m}} }$ and we have always $%
\frac{\beta(1-q_{0})-1}{\beta(1+\lambda_{\mathrm{m}})-2}>0$. Two possibles late time
behaviour of the universe can be found. The case $\lambda_{\mathrm{m}}<1-2q_0 <3$, i.e. $%
\frac{\lambda_{\mathrm{m}}+2q_{0}-1}{\beta(1+\lambda_{\mathrm{m}})-2}>0$, where the universe hits a BR
singularity in the far future. The case $1-2q_0<\lambda_{\mathrm{m}}<3$, i.e. $\frac{%
\lambda_{\mathrm{m}}+2q_{0}-1}{\beta(1+\lambda_{\mathrm{m}})-2}<0$, where the dimensionless Hubble rate
decreases and vanishes at $x_b$ {(given in Eq.~(\ref{Bounce-1})) and the universe bounces at
this point.}
In the far future, after the bounce ($x\rightarrow -\infty$), the 
Hubble rate vanishes and the universe tends to a Minkowski state.
\end{enumerate}
The late time behaviour of the  Hubble rate $E(x)$ for the case $\lambda_{\mathrm{m}}=0$, is the same as the general one given by Eq.~(\ref{sol-D3}); see table \ref{Fate-Universe} for more details.}}

\subsubsection{The case $\Delta=0$}

In this case, where $\beta =\beta _{1}$ or $\beta =\beta_{2}$, the
Hubble rate  is given by Eq.~(\ref{solution2}):
\begin{equation}
E^{2}\ =\ (\mathbf{A}_{0}+\mathbf{A}_{1}x)\exp \left( \frac{2+(\lambda _{\mathrm{m}} -\lambda
_{\mathrm{H}}-7)\beta }{2\beta }x\right),  \label{sol-D4}
\end{equation}%
where $\mathbf{A}_{0}=1$ and $\mathbf{A}_{1}$ are defined in Eq.~(\ref{A1}) as $%
\mathbf{A}_1=-\sigma_0-2(1+q_0)$, where $\sigma_0$ is the argument of the exponential
term in the equation above. 
{The total EoS of the universe in this case reads
\begin{equation}
\omega_{\rm tot}\ =\ -1-\frac{\mathbf{A}_{1}}{3(1+\mathbf{A}_{1}x)}+\frac{\sigma_0}{3} \ ,
\end{equation}
which converges to $-1+\frac{\sigma_0}{3}$ as $x\rightarrow -\infty$.}

Following the discussion in item 
\ref{item-D0} above, we consider only the case $\beta
=\beta _{2}$. %
For $\beta<\frac{2}{7+\lambda_{\mathrm{H}}-\lambda_{\mathrm{m}}}$, the
argument of the exponential term  is always positive ($\sigma_0>0$), {while  
$\mathbf{A}_{1}<0$}. This indicates that the universe undergoes a bounce at some 
$x_b=\frac{1}{|\mathbf{A}_1|}$ in the future. Hereafter, the universe starts to collapse
and as $x\rightarrow-\infty$  it converges into a Minkowski state.
For $\frac{2}{7+\lambda_{\mathrm{H}}-\lambda_{\mathrm{m}}}<\beta<
\frac{2}{1+\lambda_{\mathrm{H}}+\lambda_{\mathrm{m}}}$ 
(cf. Eq.~(\ref{beta-possible})), the argument of the exponential term is always
negative ($\sigma_0<0$): 
{if  $|\sigma_0|>2(1+q_0)$, then $\mathbf{A}_{1}>0$;   the
exponential  term decays faster than the linear term $\mathbf{A}_{1}x$, thus,  the
universe tends to a Minkowski state in the far future.
However, when  $|\sigma_0|<2(1+q_0)$ (so that $\mathbf{A}_{1}<0$),
the universe will bounce at $x_b=\frac{1}{|\mathbf{A}_1|}$ in the future, afterwards, it will start collapsing and 
in an infinite time, where $x\rightarrow-\infty$ (i.e., $a\rightarrow0$), 
the Hubble rate and  its time derivative  diverge.
Therefore, the universe meets a LB at late time. Notice that, in this case ($\sigma_0<0$),
the matter content of the universe at late time behaves as a phantom matter.}

In the case of interaction parameter $\lambda_{\mathrm{m}}=0$,  
if $\lambda_{\mathrm{H}} >9$ and ${\sigma}_{0}>-2(1+q_{0})$,
the term in the bracket in Eq. (\ref{sol-D4}) is negative, while the argument of the exponential
on the same equation is positive (for $x>0$). Therefore, the
universe bounces in the future at some $x_b=1/|\mathbf{A}_{1}|$. After the
bounce, the universe starts collapsing towards $x<0$.
Hereafter, the term in the bracket will be always positive, while the
argument of the exponential term will evolve negatively. Whence, in the far
future ($t\rightarrow +\infty$), as $x\rightarrow-\infty$ the Hubble rate
tends to zero and the universe tends to a  flat Minkowski state.

Finally, for the choice of $\lambda_{\mathrm{H}}=9$, we get ${\sigma}_{0}=0$, $%
\beta=1/8$ and $\mathbf{A}_1=-2(1+q_0)<0$. In this case, Eq.~(\ref{sol-D4}%
) reduces to Eq.~(\ref{LR1}), we have
\begin{equation}
E^{2}(x)\ =\ 4\left(4\Omega_{\mathrm{H}_{0}}-1\right)x+1.   \notag
\end{equation}
This implies a bouncing scenario for the universe at $x=x_b=\frac{1}{%
2(1+q_0)}$.
{Hereafter, the universe starts to collapse during an infinite time,
and as $x$ tends to $-\infty$ the scale factor tends to zero, the Hubble rate diverges
while its time derivative remains finite; $\dot{E}=2H_0(4\Omega_{\mathrm{H}_{0}}-1)$. Therefore, in the far future ($t\rightarrow +\infty$), the universe hits  a {LSBB} abrupt event.}
In summary, what happens is that the universe evolves from a LSBB in the past, expands until it bounces and heads back to a LSBB.
Whenever  $\mathbf{A}_{1}<0$,   the dimensionless parameter
$\Omega_{\mathrm{H}_0}$ satisfies $\Omega_{\mathrm{H}_0}<\frac{1}{4}$.
Consequently, by substituting $\beta=1/8$
in the Eq.~(\ref{EoS2}) we find that the EoS  lies in the range
\begin{eqnarray}
-\frac{16}{3}<\omega_{\mathrm{H}}<-4~,   \quad \quad (0<t<+\infty). \notag
\end{eqnarray}
This relation implies that, the universe starts from a LSBB in the past with EoS  for the HRDE $\omega_{\mathrm{H}}=-4$, and reaches a  bounce when
$\omega_{\mathrm{H}}=-\frac{16}{3}$ at some $t_b$ in the future. After the bounce, the universe will collapse and will  hit  a LSBB abrupt event
in the far future ($t\rightarrow+\infty$) with the same EoS,  $\omega_{\mathrm{H}}=-4$, of the initial state.
Since the EoS  of the dark energy evolves in the
region $\omega_{\mathrm{H}}<-1$,  the corresponding interacting
HRDE model represents a phantom-like behaviour, although its EoS
is too negative to describe nowadays universe.
Notice that, when the  universe
hits a LSBB  in the far future,  the total EoS, $\omega_{\mathrm{tot}}$, of  the
universe tends to $-1$ which has the same value for the total EoS at the initial LSBB abrupt event.

If $\beta=\frac{2}{7+\lambda_{\mathrm{H}}-\lambda_{\mathrm{m}}}$, for the
particular case $\beta =\frac{2}{7}$, the interaction parameters are equal and the condition
(\ref{caseSolB}) is satisfied for $\lambda _{\mathrm{H}}=\lambda_{\mathrm{m}}=\frac{9}{7}$. Then, the Hubble rate
is obtained from Eq.~(\ref{LR1}) as
\begin{equation}
E^{2}(x)\ =\ 2\left( \frac{7}{2}\Omega _{\mathrm{H}_{0}}-2\right) x+1.
\label{LR1-2b}
\end{equation}%
{On the case where the holographic dimensionless parameter
$\Omega_{\mathrm{H}_{0}}$ satisfies $\Omega _{\mathrm{H}_{0}}<4/7$,
the universe undergoes a \emph{bounce} followed by a  LB   in the far  future.
It can be shown that at the LB $\omega_{\mathrm{H}}(a\rightarrow-\infty)=-\frac{7}{4}$, whereas at the bounce
it reads $\omega_{\mathrm{H}}(x=\frac{1}{|A_1|})=-\frac{7}{3}$. Therefore, $\omega_{\mathrm{H}}<-1$ and the corresponding interacting HRDE model has a phantom-like behaviour. Moreover, when the universe hits a  LB    in the far future,
the EoS of HRDE tends to $-\frac{7}{4}$ again.}
In fact, there is a symmetric evolution with respect to the source; i.e. the universe heads from a LB to a bounce and then back to a LB.
All the  solutions we have analysed in section \ref{Sing-model}  are summarised in table \ref{Fate-Universe}.


\begin{center}
\begin{table*}[t!]
  \centering
  \begin{tabular}{ccccccc}
    \toprule
    \hline\hline
 Sec. & Interacting    &  ~  $\Delta$~ & $\beta$ & $\lambda$ & $\sigma_0, \sigma_{\pm}$ &    Late time  \\
    &   model  & &&  & &   behaviour \\ \hline\hline
    \midrule
  \ref{Sing-model1} &            $\lambda_{\rm m}=\lambda_{\rm H}=0$ & {$\Delta>0$} &  $\beta=\frac{1}{2}$ & -- & $\sigma_-=-3$,~ $\sigma_+=0$ &   de Sitter  \\
  &        & {$\Delta>0$} &   $\beta<\frac{1}{2}$ & -- & $\sigma_-=-3$,~ $\sigma_+>0$   &     BR  \\
\hline
\ref{subsec.B} &             $\lambda_{\rm m}=8\beta -1$,   & $\Delta=0$ &   $\beta<\frac{1}{2}$ & ${\lambda_{\rm c}=0}$ & $\sigma_0=0$  &     LSBB  \\
      &              $\lambda_{\rm H}=\frac{2}{\beta}(2\beta-1)^2$  & & &{$\lambda_{\rm c}\neq 0$}& &   {LR} \\ \hline
\ref{Sing-model3}  &     $\lambda_{\rm m}\neq 0$,  $\lambda_{\rm H}=0$ & $\Delta=0$ &   $\beta<\frac{1}{2}$ & -- & $\sigma_0>0$ &    Minkowski   \\
 &      & $\Delta>0$ &  $\beta<\frac{2}{1+\lambda_{\rm m}}\leq \frac{1}{2}$ &  $\lambda_{\rm m}\geq3$ & $\sigma_\pm>0$,  &    Minkowski \\
    &   & $\Delta>0$ &  $\beta<\frac{1}{2}<\frac{2}{1+\lambda_{\rm m}}$ &  $\lambda_{\rm m}<1-2q_0<3$ & {$\sigma_{-}<0$},~ {$\sigma_+>0$} &     BR  \\
&       & $\Delta>0$ &  $\beta<\frac{1}{2}<\frac{2}{1+\lambda_{\rm m}}$ &  $1-2q_0<\lambda_{\rm m}<3$ &
      {$\sigma_{-}<0$},~ {$\sigma_+>0$} &      Minkowski   \\
      \hline
 \ref{Sing-model3} &    $\lambda_{\rm m}=0$, $\lambda_{\rm H}\neq 0$ & $\Delta=0$ &  $\beta<\frac{2}{1+\lambda_{\rm H}}$  &  $\lambda_{\rm H}<9$ &  {${\sigma}_0<-2(1+q_0)<0$} &    Minkowski  \\
    &   & $\Delta=0$ &  $\beta<\frac{2}{1+\lambda_{\rm H}}$  &  $\lambda_{\rm H}<9$ &   {$-2(1+q_0)<{\sigma}_0<0$} &    LB  \\
  &             & $\Delta=0$ &  $\beta<\frac{2}{1+\lambda_{\rm H}}$ &  $\lambda_{\rm H}>9$ &   {$-2(1+q_0)<0<{\sigma}_0$}  &     Minkowski   \\
   &       & $\Delta=0$ &  $\beta=\frac{1}{8}$  &  $\lambda_{\rm H}=9$ & ${\sigma}_0=0$  &     LSBB\footnote{This is a specific case of  the solution of Eq.~(\ref{caseSolB}) given in the section \ref{subsec.B} (cf. see the third row in the table above).}   \\
    &    &  $\Delta>0$ &  $\beta<\frac{2}{7+\lambda_{\rm H}}$ &  $\lambda_{\rm H}>3$ & ${\sigma}_+>0$,  &     BR   \\
     &     &   &  &  & {${\sigma}_-<0$}  &      \\
    &          &  $\Delta>0$ &   $\frac{2}{7+\lambda_{\rm H}}<\beta<\frac{3}{2(3+\lambda_{\rm H})}$ &   $\lambda_{\rm H}>0$ & ${\sigma}_+>0$,   &    BR   \\
      &          &  &   &  & {${\sigma}_-<0$}&       \\
   &     & $\Delta>0$ &   $\frac{3}{2(3+\lambda_{\rm H})}<\beta<\frac{1}{2}$ &  all $\lambda_{\rm H}$ &{${\sigma}_\pm<0$}  &     Minkowski   \\
    &    & $\Delta>0$  &  $\beta=\frac{3}{2(3+\lambda_{\rm H})}<\frac{1}{2}$ &  all $\lambda_{\rm H}$ & ${\sigma}_+=0$,  &      de Sitter   \\
 &            &  &  &  & {${\sigma}_-<0$}  &        \\
     \hline
 \ref{Sing-model3} &    $\lambda_{\rm m}\neq0$,   $\lambda_{\rm H}\neq0$   & $\Delta=0$ &  $\beta=\beta_2<\frac{2}{7+\lambda_{\rm H}-\lambda_{\rm m}}$  &  --  & $\sigma_0>0$  &  Minkowski  \\
 &      & $\Delta=0$ &  $\frac{2}{7+\lambda_{\rm H}-\lambda_{\rm m}}<\beta=\beta_2<\frac{2}{1+\lambda_{\rm m}+\lambda_{\rm H}}$  & -- &
     {${\sigma}_0<-2(1+q_0)<0$}  &     Minkowski   \\
  &        & $\Delta=0$  &  $\frac{2}{7+\lambda_{\rm H}-\lambda_{\rm m}}<\beta=\beta_2<\frac{2}{1+\lambda_{\rm m}+\lambda_{\rm H}}$ & -- &  {$-2(1+q_0)<{\sigma}_0<0$}  &     LB   \\
   &              & $\Delta>0$ &  $\beta<\frac{2}{7+\lambda_{\rm H}-\lambda_{\rm m}}$ or  & -- & {$\sigma_+>0$}   &     BR    \\
 &  &  &  $\frac{2}{7+\lambda_{\rm H}-\lambda_{\rm m}}<\beta<\frac{3-\lambda_{\rm m}}{6+2(\lambda_{\rm H}-\lambda_{\rm m})}$ &  & {$\sigma_-<0$}   &        \\
 &         & $\Delta>0$  &  $\beta>\frac{3-\lambda_{\rm m}}{6+2(\lambda_{\rm H}-\lambda_{\rm m})}$  &  -- & {${\sigma}_\pm<0$}   &   Minkowski  \\
&          & $\Delta>0$ &  $\beta=\frac{3-\lambda_{\rm m}}{6+2(\lambda_{\rm H}-\lambda_{\rm m})}$ & -- & $\sigma_+=0$, {${\sigma}_-<0$} &      de Sitter   \\
                  \hline\hline
    \bottomrule
  \end{tabular}
  \caption{Summary of the behaviours of the universe at late times, for the physical range of holographic parameters $\beta<\frac{1}{2}$, for different DM and DE interactions.}
  \label{Fate-Universe}
\end{table*}
\end{center}

\section{Summary and Conclusions}

In this paper, we have considered a HRDE \cite{Holographich}, as the dark energy component of the
universe, coupled to the CDM component of the universe.
As it is well known, in the absence of  interaction ($Q=0$) and
depending on the physically relevant range of the holographic parameter $\beta$,
the late time universe will  end up in a big rip  (BR) singularity if $\beta<\frac{1}{2}$  \cite{Holographich}. We remind as well that those values of $\beta$ are consistent with the latest observations.

On this work, we are interested in investigating whether or not different
interactions between HRDE and CDM components of the universe 
could resolve or {smoothen} the BR singularity.
We considered different interaction functions  such as
$Q=\lambda_{\mathrm{m}} H\rho_{\mathrm{m}}$, $Q=\lambda_{\mathrm{H}} H\rho_{\mathrm{H}}$ and
$Q= H(r\lambda_{\mathrm{m}}+\lambda_{\mathrm{H}})\rho_{\mathrm{H}}+{\lambda_{\mathrm{c}} H\rho_{\mathrm{c}}}$ to
study the late time behaviour of the universe ($r$ is defined in Eq.~(\ref{r})).
In the presence of the general function $Q= H(r\lambda_{\mathrm{m}}+\lambda_{\mathrm{H}})\rho_{\mathrm{H}}+{\lambda_{\mathrm{c}} H\rho_{\mathrm{c}}}$,  the corresponding differential equation (\ref{E second}) governing the dynamical evolution  of the universe, possesses two types of generic solutions (cf. Eqs.~(\ref{solution}) and (\ref{solution2})) depending on the  discriminant function $\Delta$, given by Eq.~(\ref{delta}).

For  different choices of the interaction constants
$\lambda_{\rm m}$, $\lambda_{\rm H}$, $\lambda_{\rm c}=0$, and the HRDE parameter $\beta$,  the physically relevant  solutions corresponding  to $\Delta>0$ (cf. table \ref{Fate-Universe}) lead  to an asymptotic behaviour of the universe with: (i) a BR singularity,  (ii) a Minkowski  or (iii) a {\rm de Sitter} state in the far future (see table \ref{Fate-Universe} for a summary of the late time behaviour of the different solutions).
In particular, we have shown that for proper combinations of $\lambda_{\rm H}$, $\lambda_{\rm m}$ and $\beta$, the BR singularity can be removed.

{On the other hand, when $\Delta=0$ (and $\lambda_{\rm c}=0$), the  general Hubble rate is given by  Eq.~(\ref{solution2}).
When $\sigma_0<0$ and $\mathbf{A}_1>0$ (or
$\sigma_0>0$ and $\mathbf{A}_1<0$), 
the universe approaches  a Minkowski state, asymptotically
(cf. table \ref{Fate-Universe}). 
However, the case  with $\mathbf{A}_1<0$ where  $\sigma_0\leq0$,   represents  two new abrupt events in the far  future of the universe. We  summarise carefully these two cases:}
\begin{enumerate}[label=(\roman*)]
\item  For  $\sigma_0=0$, the Hubble rate reads $E^2(x) =1- |\mathbf{A}_1|x$, while $\dot{E}=const.$. This solution  bounces  at $x_b=\frac{1}{|\mathbf{A}_1|}$. After this point, the universe starts to collapse and as $x\rightarrow-\infty$, the Hubble rate diverges while its time derivative remains finite.
This corresponds to a new abrupt event, which is smoother than the big bang singularity and happens in  an infinite cosmic  time. We have called it  the ``little sibling of the big bang'' (LSBB).
\item
For $\sigma_0<0$, the Hubble rate reads $E^2 = \ (1- |\mathbf{A}_1|x)e^{-|\sigma_0|x}$, therefore $\dot{E} =  -\frac{H_0}{2}[|\mathbf{A}_1|+|\sigma_0|(1- |\mathbf{A}_1|x)]e^{-|\sigma_0|x}$. This solution  indicates that, the universe bounces first at $x_b=\frac{1}{|\mathbf{A}_1|}$, then will collapse, and as $x\rightarrow-\infty$, the Hubble rate and its time derivative diverge at an infinite time.
Therefore,  in the far future ($t\rightarrow+\infty$), the universe tends to another type of abrupt event which we  have called the ``little bang" (LB) (cf. table \ref{Fate-Universe}).
\end{enumerate}
{{Finally, a similar analysis for  the case $\Delta=0$ and $\sigma_0=0$, for a non-vanishing interaction parameters $\lambda_{\rm c}$, represented a  universe which starts  its evolution from an infinite past with a LB, and afterwards it expands until it  bounces at a finite time in the future, and then it recollapses again  to a LB; or it can start from a bounce and heads to a little rip at an infinite cosmic  time.
For  $\lambda_{\rm c}>0$, there is a unique Lorentzian solution that interpolate between two bounces.
}}

These new types  of abrupt events arise  only when an interaction between HRDE and CDM is present. We have further  shown that,  under these conditions  the HRDE may have a phantom like behaviour.

\section*{Acknowledgements}

The work of MBL is supported by the Basque Foundation of Science IKERBASQUE. She also wishes to acknowledge the partial support from the Basque government Grant No. IT956-16 (Spain) and FONDOS FEDER under grant FIS2014-57956-P (Spanish government).
The work of YT was supported by the Polish Narodowe Centrum Nauki (NCN) grant 2012/05/E/ST2/03308. He also acknowledges the Bonyad-e-Melli Nokhbegan of Iran (INEF) and the Brazilian agencies FAPES/CAPES for partial ﬁnancial supports.
This article is based upon work from COST Action CA15117 “Cosmology and Astrophysics Network for Theoretical Advances and Training Actions (CANTATA)”, supported by COST (European Cooperation in Science and Technology).



\end{document}